\documentclass{ifacconf}

\usepackage{graphicx}      
\usepackage{natbib}        
\usepackage{amsmath, amssymb}
\usepackage{multirow}
\usepackage{multicol}
\usepackage{subcaption}
\usepackage{float}
\usepackage{cuted}
\usepackage{xcolor}
\usepackage{url}

\DeclareMathOperator{\sign}{sgn}

\begin{document}
\begin{frontmatter}

\title{Fuel-Optimal Powered Descent Guidance for Hazardous Terrain} 


\author[First]{Sheikh Zeeshan Basar} 
\author[Second]{Satadal Ghosh} 

\address[First]{Department of Aerospace Engineering, Indian Institute of Technology, Madras, India (e-mail: ae21s013@smail.iitm.ac.in).}
\address[Second]{Room 222, CTC Building, Department of Aerospace Engineering, Indian Institute of Technology, Madras, India (e-mail: satadal@iitm.ac.in)}

\begin{abstract}                
Future interplanetary missions will carry more and more sensitive equipment critical for setting up bases for crewed missions. The ability to manoeuvre around hazardous terrain thus becomes a critical mission aspect. However, large diverts and manoeuvres consume a significant amount of fuel, leading to less fuel remaining for emergencies or return missions. Thus, requiring more fuel to be carried onboard. This work presents fuel-optimal guidance to avoid hazardous terrain and safely land at the desired location. We approximate the hazardous terrain as step-shaped polygons and define barriers around the terrain. Using an augmented cost functional, fuel-optimal guidance command, which avoids the terrain, is derived. The results are validated using computer simulations and tested against many initial conditions to prove their effectiveness.
\end{abstract}

\begin{keyword}
Aerospace, Space exploration and transportation, Guidance, navigation and control of vehicles, Autonomous systems
\end{keyword}

\end{frontmatter}

\section{Introduction}
While several moon landing missions like Luna 2 of the erstwhile-Soviet Union and Rangers of the NASA (U.S.) have been directed toward an intentionally controlled crash landing (lunar impact) for certain specific reasons (\cite{nasa}) most of the spacecraft landing missions, especially which are required to function and observe several parameters of interest over a long period of time, require a soft landing on celestial bodies for ensuring the survival of the spacecraft upon landing. A typical soft landing on a rocky planet has several stages. Among them, during the powered descent stage, the guidance law must be robust to perturbation, consume the least fuel, and land softly as close to the desired site as possible. The study of autonomous guidance laws for powered descent began with the Apollo era. However, most guidance laws were rudimentary, as they were severely limited by the computational power and memory. In fact, the Apollo missions used a straightforward polynomial-based guidance law (\cite{Klumpp_1974}). 

The polynomial-based guidance is relatively easier to implement; however, they lack in fuel-optimality, robustness and the required accuracy. Feedback guidance laws have been presented in the literature to alleviate the above-mentioned issues. Classical feedback laws for missile guidance, like Proportional Navigation (PNG) as discussed in \cite{Zarchan_2012}, have inspired some of the earliest guidance laws like biased-PNG (\cite{Byung_Jang_Hyung_1998}) and pulsed-PNG (\cite{Wie_1998}) for powered descent. Extending the idea of zero-effort-miss (ZEM) from missile guidance laws \cite{Ebrahimi_Bahrami_Roshanian_2008} presented a fuel-optimal spacecraft landing guidance law, where another notion of zero-effort-velocity (ZEV) was also included. The ZEM/ZEV-based optimal powered descent guidance was made robust against perturbations using Sliding Mode Control (SMC), and its variants (\cite{Furfaro_2011}, \cite{Furfaro_Gaudet_Wibben_Kidd_Simo_2013}, \cite{Wibben_Furfaro_2016}, \cite{Furfaro_Cersosimo_Wibben_2013}). Missile guidance concepts, such as collision course and heading error, were further incorporated by \cite{VS_Ghosh_2021} to develop a novel SMC-based guidance law in full nonlinear setting, which was further extended in \cite{Shincy_Ghosh_2022} to include a notion of fuel optimality as well.

A critical drawback of the optimal guidance law is that a part of the trajectory maybe underground for a large terminal time implying a crash landing at an undesired location, thus defeating the purpose of the mission. While lowering terminal time may help in avoiding this issue, it requires a significantly larger thrust command to succeed. \cite{Guo_CTVG_2012} suggested use of waypoints to avoid crashing into the planetary surface, while selection of such waypoints optimally was studied by \cite{Guo_Hawkins_Wie_2013}. However, the method developed there becomes computationally expensive as the number of time-steps increases and if both terminal and waypoint times are left free for the optimiser to solve.

In order to avoid collision during the powered descent phase, an improved ZEM/ZEV feedback guidance was presented in \cite{Zhou_Xia_2014} by including a switching-form term as penalty in the performance index. The limitation posed by dependence on prior experience in the added term therein was subsequently obviated in \cite{Zhang_Guo_Ma_Zeng_2017} by using a self-adjusting augmentation to the performance index. Most of the existing literature on spacecraft landing while also avoiding collision consider the terrain around the landing site to be relatively flat, which is usually addressed by using a simple glideslope constraints to avoid minor undulation and debris near the surface. To attempt landing at treacherous terrain, glideslope constraints may be too conservative. In this context, terrain avoidance guidance was presented by \cite{Gong_Guo_Ma_Zhang_Guo_2021}, where Barrier Lyapunov functions were used to avoid crashing into the terrain. In another study by \cite{Gong_Guo_Ma_Zhang_Guo_2022} terrain avoidance was achieved by using prescribed performance functions. Both of these guidance laws were able to avoid crashing into terrain and safely land. However, the guidance law developed requires rate information of the intermediate control, which is parameterised by several time-dependent variables, to be estimated. Further, these methods also lacked in fuel optimality and satisfactory performance under thrust constraints.

To this end, an Optimal Terrain Avoidance Landing Guidance (OTALG) is presented in this paper to land safely at the desired site, while avoiding uneven terrain and maintaining near-fuel optimality. The terrain is approximated as $n$ step-shaped polygons using readily available terrain data. Based on the step-dimensions of the polygons, We then introduce barrier polynomials as a boundary function to cover the terrain. The barrier function, which is a collection of these polynomials, is assumed to be stored \textit{a-priori} on-board the spacecraft. A novel augmentation to the performance index is introduced next, which introduces penalty if the lander moves close to the barriers. Fuel optimal guidance law is developed as a function of time dependent ZEM and ZEV, both of which can be calculated online, given the lander's current position and velocity, time-to-go and desired terminal conditions. The efficacy of our work in terrain avoidance shown using computer simulations. Using Monte-Carlo simulations, we demonstrate near-fuel optimality of the proposed guidance law. To the best of author's knowledge, the proposed guidance law is the first attempt at terrain avoidance guidance law for powered descent that also ensures near-fuel-optimality.

The rest of the paper is organised as follows. Section \ref{sec:Problem_Preliminaries} details the landing dynamics and some preliminaries from the optimal control-based guidance laws. Then using piecewise-smooth polynomials and information about the terrain, barriers are developed in Section \ref{sec:BarrierDefn}. Section \ref{sec:GuideDev} introduces the terrain avoidance performance index, and the optimal guidance law is derived subsequently. Section \ref{sec:Sims} presents the simulation results and observations. Finally, the paper is concluded in Section \ref{sec:Concl}. 

\section{Problem Formulation and Preliminaries}\label{sec:Problem_Preliminaries}

\begin{figure}
    \centering
    \includegraphics[width = 0.8\linewidth]{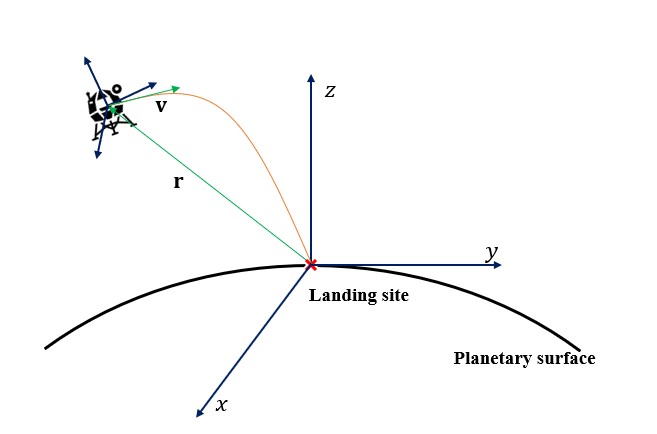}
    \caption{Spacecraft Landing Geometry}
    \label{fig:dynamics}
\end{figure}

The non-rotating inertial ENU-frame with origin at the landing site is considered as shown in Fig. \ref{fig:dynamics}. Assuming a 3-DOF dynamics, the lander can be modelled as:
\begin{equation} \label{eq: dynamics}
\begin{array}{ll}
    \dot{\mathbf{r}} &= \mathbf{v}\\
    \dot{\mathbf{v}} &= \mathbf{a} + \mathbf{g} + \mathbf{a}_\mathrm{p}\\
    {\mathbf{a}} &= \frac{\mathbf{T}}{m}\\
    \dot{m} &= -\frac{\Vert \mathbf{T} \Vert}{I_{\mathrm{sp}}g_e}
\end{array}
\end{equation}
where, $\mathbf{r} = [r_x, r_y, r_z]^\mathrm{T}$ and $\mathbf{v} = [v_x, v_y, v_z]^\mathrm{T}$ represent the position and velocity of spacecraft, and $\mathbf{g} = [g_x, g_y, g_z]^\mathrm{T}$ is the local gravity. The guidance command provided by the thrusters is represented by $\mathbf{a} = [a_x, a_y, a_z]^\mathrm{T}$, $\mathbf{a}_p$ is the net acceleration caused due to bounded perturbations (e.g. wind), $m$ is the lander's mass, $I_{\mathrm{sp}}$ is the specific impulse, and $g_e$ is the gravitational acceleration of Earth. Since the altitude at which the powered descent stage starts is significantly smaller than the radius of planets, it is valid to assume a constant gravitational acceleration acting only along the z-axis, that is, $\mathbf{g} = [0,0,-g]^\mathrm{T}$.

The core idea of ZEM and ZEV is to determine the deviation from the desired position ($\mathbf{r}_f$) and velocity ($\mathbf{v}_f$) at the final time $t_f$ if no control effort is applied from current time $t$. Defining time-to-go, $t_{\mathrm{go}} \triangleq t_f - t$, we have:
\begin{equation} \label{eq:zeroeffort}
    \begin{array}{ll}
         \mathbf{ZEM}(t) &= \mathbf{r}_f - \left[\mathbf{r}(t) + \mathbf{v}t_{\mathrm{go}} + \frac{1}{2}\mathbf{g}t_{\mathrm{go}}^2\right]  \\
         \mathbf{ZEV}(t) &= \mathbf{v}_f - \left[ \mathbf{v}(t) + \mathbf{g}t_{\mathrm{go}} \right]
    \end{array}
\end{equation}
where, $\mathbf{r}_f=\mathbf{r}(t_f)$, and $\mathbf{v}_f=\mathbf{v}(t_f)$ represent the position and velocity at the final time $(t_f)$. The classical ZEM/ZEV-based optimal guidance law of \cite{Ebrahimi_Bahrami_Roshanian_2008} was derived by considering the performance index,
\begin{equation} \label{eq:OGperfidx}
    J = \frac{1}{2} \int_{t_0}^{t_f} \mathbf{a}^\mathrm{T}\mathbf{a}\,d\mathrm{t},
\end{equation}
subjected to the dynamics in (\ref{eq: dynamics}), with $\mathbf{a}_\mathrm{p} = 0$. The optimal guidance law was obtained there as,
\begin{equation} \label{eq:classical_acc}
    \mathbf{a} = \frac{6}{t_{\mathrm{go}}^2}\mathbf{ZEM}(t) - \frac{2}{t_{\mathrm{go}}}\mathbf{ZEV}(t).
\end{equation}

\section{Barrier definition around hazardous terrain}\label{sec:BarrierDefn}

\begin{figure}
\begin{center}
\includegraphics[width=0.8\linewidth]{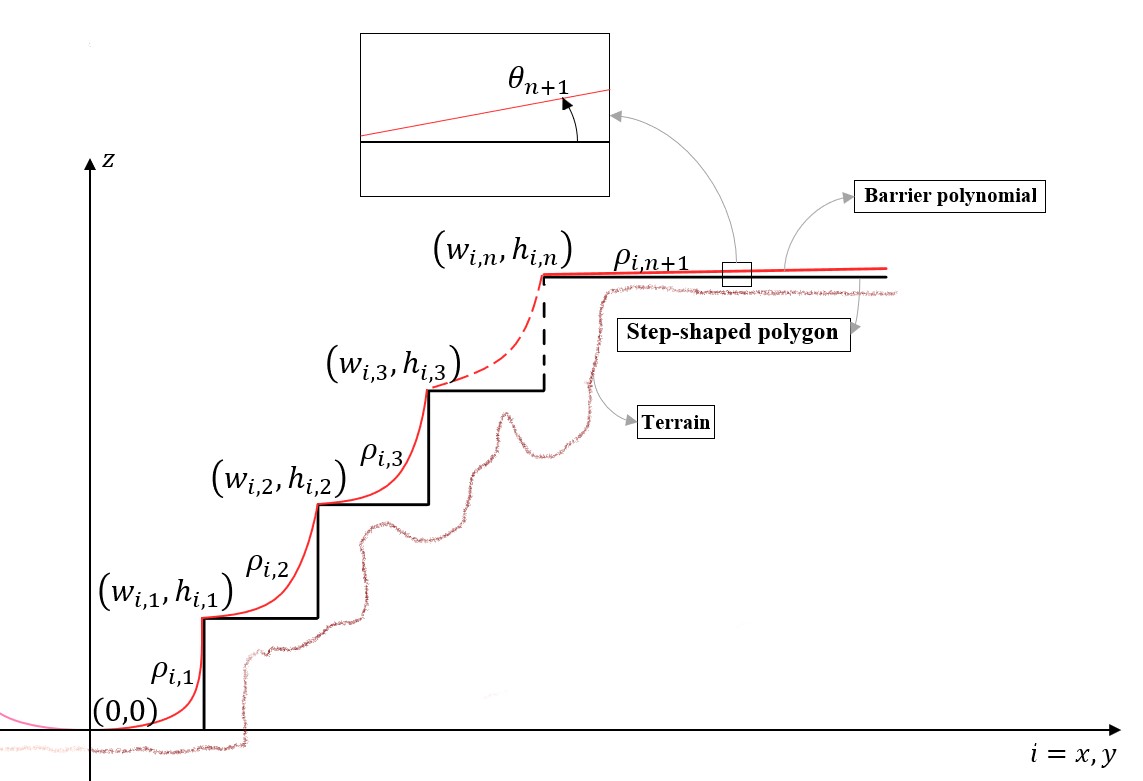}
\caption{Illustration of terrain and barrier around $n$-step shaped polygons.}
\label{fig:barrier}
\end{center}
\end{figure}

Terrain data available from reconnaissance steps/missions are considered to be used to create the $n$ number of step-shaped polygons to approximate the terrain, as shown in Fig. \ref{fig:barrier}. Defining the height of $j^\textsuperscript{th}$ step as $h_{i,j}$, and the horizontal distance from the landing site (origin) along $i$-axis as $w_{i,j}$, where $i = x,\, y$. 

\subsection{Barriers for horizontal motion}\label{subsec:BarrierHorz}

To restrict the spacecraft motion in the horizontal planes, we define a barrier polynomial $\rho_{i,j}$ as in (\ref{eq:rho_lat}). For $n$-steps, we will define $n+1$ number of barriers, where the order of the polynomial is determined on the basis of degree of conservatism the mission demands. A higher degree polynomial will closely follow the step shape. Allowing for more freedom of movement for the lander. We further consider the $(n+1)^\textsuperscript{th}$ barrier to be a linear polynomial instead of any higher order polynomial. 
\begin{equation} \label{eq:rho_lat}
    \rho_{i,j} = \left\{\begin{array}{ll}
         &\pm \left( \beta_{i,j}(r_z + \gamma_{i,j})^{\frac{1}{\lambda_{i,j}}} + \alpha_{i,j}\right),\,h_{i,(j-1)} \leq r_z \leq h_{i,j}\\
         &\pm \left( \beta_{i,(n+1)}(r_z + \gamma_{i,(n+1)}) + \alpha_{i,(n+1)}\right),\,r_z \geq h_{i,n}
    \end{array}\right.
\end{equation}

where $i = x,\,y$ and $j = 1,\, \dots,\, n$.
From the first barrier to the $n^\textsuperscript{th}$ barriers, the constants are defined as:
\begin{align} \label{eq: barrier_j}
    \alpha_{i,j} = w_{i,(j-1)};\:\:
    \beta_{i,j}= \frac{w_{i,j} - w_{i,(j-1)}}{\left(h_{i,j} - h_{i,(j-1)}\right)^{\frac{1}{\lambda_{i,j}}}};\:\:
    \gamma_{i,j} = -h_{i,(j-1)}
\end{align}

and $\lambda_{i,j}$ is a positive, even natural number. Further note that, $h_{i,0} = 0$, and $w_{i,0} = 0$, that is, origin is the landing site. For the $(n+1)^\textsuperscript{th}$ barrier, we first choose the slope angle of the barrier, $\theta_{(n+1)}$ with respect to the axis under consideration. The angle can be chosen as a small value (approx. $0.05^{\circ}\,-\,0.1^{\circ}$) or a high value for a relatively flat plateau or a hill, respectively, beyond the canyon containing the lading site. The constants can now be defined as:
\begin{align} \label{eq: barrier_n1}
    \alpha_{i,(n+1)} &= w_{i,n};\:\:
    \gamma_{i,(n+1)} = -h_{i,n};\nonumber\\
    \beta_{i,(n+1)} &= \tan{\left({\pi}/{2} - \theta_{(n+1)} \right)}
\end{align}

\subsection{Barriers for vertical motion}\label{subsec:BarrierVert}

It is imperative to define barriers at specific heights to avoid crash landing into the terrain due to vertical motion. We do this by including a small margin, $\delta$, to the height of the next lower barrier. The designer can choose this margin of safety as per requirement. However, here we run into a problem that if the lander is within the lateral bound of $j^\textsuperscript{th}$ step, but is above the height of $(j+1)^\textsuperscript{th}$ step, the lander would keep on bouncing off the vertical barrier corresponding to the $(j+1)^\textsuperscript{th}$ step. To address this issue, we select the vertical barrier using the following simple comparison:

\begin{equation}\label{eq:rho_z}
    \rho_{z,j} = \left\{\begin{array}{ll}
    h_{i,n} + \delta, & r_z \geq h_{i,n}\\
    h_{i,(j-1)} + \delta, & \left( h_{i,(j-1)} \leq r_z \leq h_{i,j} \right)\ \mathrm{AND}\\
                  &\left( w_{i,(j-1)} \leq \Vert [r_x,\, r_y]^\mathrm{T} \Vert_{\infty} \leq w_{i,j} \right)\\
    h_{i,j} + \delta, & \left( h_{i,(j-1)} \leq r_z \leq h_{i,j} \right)\ \mathrm{AND}\\
                  &\left( w_{i,(j-1)} \geq \Vert [r_x,\, r_y]^\mathrm{T} \Vert_{\infty} \geq w_{i,j} \right)
    \end{array}\right.
\end{equation}
where, $j = 1,\,\dots,\,n$.

\section{Development of feedback guidance}\label{sec:GuideDev}

In this section, a novel feedback guidance law is developed, which can navigate the terrain and land safely and softly at the desired landing site consuming least amount of fuel.

\subsection{Performance index for collision avoidance}\label{subsec:PerfIdx}

To avoid crashing into the local terrain, we introduce a novel augmentation to the performance index.
\begin{equation} \label{eq:newJ}
    J = \frac{1}{2}\int_{0}^{t_f} \left[ \mathbf{a}^\mathrm{T}\mathbf{a} - \sum_{i} l_{3,i}e^{-\phi_i} \right]\, d\mathrm{\tau}
\end{equation}
where, $i\,=\,x,\,y,\,z$, $l_{3,i} > 0$ is a constant, and $e^{-\phi_i}$ is the augmentation term, where $\phi_i$ is defined as:
\begin{equation}
    \phi_i = {l_{2,i}}/{(d_i^2 + l_{1,i})} \label{eq:phi_i}
\end{equation}
where, $d_i \triangleq r_i - \rho_{i,j}$. Physically, $d_i$ represents how far the lander is with respect to the barrier surface. When the lander is far away from the barriers, the augmentation is large and positive, so the term inside the integration in (\ref{eq:newJ}) is small. When the lander is close to the barrier, the augmentation is almost zero, increasing the cost, thus generating an acceleration command in direction opposite to the direction of motion to avoid crashing into terrain.

\subsection{Guidance law}\label{subsec:Law}

With the performance index in (\ref{eq:newJ}) subject to the dynamics in (\ref{eq: dynamics}) with $\mathbf{a}_\mathrm{p} = 0$, the Hamiltonian is written as:
\begin{equation} \label{eq:Hamiltonian}
    H = \frac{1}{2} \left[ \mathbf{a}^\mathrm{T}\mathbf{a} - \sum_{i} l_{3,i}e^{-\phi_i} \right] + \mathbf{p}_r^\mathrm{T}\mathbf{v} + \mathbf{p}_v^\mathrm{T}\left(\mathbf{a} + \mathbf{g}\right).
\end{equation}
Since (\ref{eq:Hamiltonian}) is not linear in $\mathbf{a}$, the optimal control is of the form:
\begin{equation} \label{eq:dHda}
    \frac{\partial H}{\partial \mathbf{a}} = 0\, \Rightarrow\, \mathbf{a} = -\mathbf{p}_v.
\end{equation}
For the dynamics (\ref{eq: dynamics}) and $\phi_i$ from (\ref{eq:phi_i}), the co-states are:
\begin{align}
    \dot{\mathbf{p}}_r &= \frac{\partial}{\partial \mathbf{r}}\left[ \frac{1}{2}\sum_{i} l_{3,i}e^{-\phi_i} \right] \label{eq:pr_dot} \\
    \dot{\mathbf{p}}_v &= -\mathbf{p}_r \label{eq:pv_dot}
\end{align}
From (\ref{eq:pr_dot}), we define $\mathbf{p} \triangleq [\dot{p}_{r_x},\, \dot{p}_{r_y},\, \dot{p}_{r_z}]$ where:
\begin{equation}
    \dot{p}_{r_i} = l_{2,i}l_{3,i}\frac{d_i\mathrm{e}^{-\phi_i}}{\left( d_i^2 + l_{1,i} \right)^2}. \label{eq: p_i}
\end{equation}
Before moving forward we note:
\begin{equation}
    \Ddot{p}_{ri} = \frac{l_{2,i}l_{3,i}}{(d_i^2 + l_{1,i})^2} \left[ 1 - \frac{4d_i^2}{(d_i^2 + l_{1,i})} + \frac{2l_{2,i}d_i^2}{(d_i^2 + l_{1,i})^2} \right]e^{-\phi_i} v_x, \label{eq: pri_ddot}
\end{equation}
and by suitably choosing $l_{1,i}$ and $l_{2,i}$ we can say that $\sign \Ddot{p}_{ri} = -\sign v_i$. Note that in horizontal motion, the lander moving close to the barriers as defined in Section \ref{subsec:BarrierHorz} can follow either of the two cases: (1) $r_i > 0$ and $v_i > 0$, or (2) $r_i < 0$ and $v_i < 0$. In the first case, from \eqref{eq: pri_ddot} and \eqref{eq:pr_dot}, $\Dot{p}_{ri} < 0$ and $\Ddot{p}_{ri} \leq 0$, which implies that $\dot{p}_{ri}$ decreases, and thus, $p_{ri} \geq p_{rfi} - t_{\mathrm{go}}\Dot{p}_{ri}$. Then, from (\ref{eq:dHda}) and (\ref{eq:pv_dot}), in the first case,
\begin{equation}
    a_i \leq -p_{vfi} - p_{rfi}t_{go} + \frac{t_{\mathrm{go}}^2}{2}\Dot{p}_{ri}. \label{eq: ai_maxQ1}
\end{equation}
On the other hand, in the second case, we have $\Dot{p}_{ri} > 0$ and $\Ddot{p}_{ri} \geq 0$ implying that $\dot{p}_{ri}$ increases, and thus, $p_{ri} \leq p_{rfi} - t_{\mathrm{go}}\Dot{p}_{ri}$. Hence, in the second case,
\begin{equation}
    a_i \geq -p_{vfi} - p_{rfi}t_{\mathrm{go}} + \frac{t_{\mathrm{go}}^2}{2}\Dot{p}_{ri}. \label{eq: ai_maxQ2}
\end{equation}

For obtaining a simplified closed form expression of $\mathbf{a}$ by simplifying (\ref{eq:dHda})-(\ref{eq:pv_dot}), an approximation $p_{ri}\approx p_{rfi} - t_{\mathrm{go}}\Dot{p}_{ri}$ is considered in this paper. Similarly we have $v_z < 0$, which implies, $p_{rz} \leq p_{rfz} - t_{\mathrm{go}}\Dot{p}_{rz}$. Following a similar analysis we get $a_z \geq -p_{vfz} - p_{rfz}t_{\mathrm{go}} + \Dot{p}_{rz}(t_{\mathrm{go}}^2/2)$, which leads to a sub-optimal guidance command as,
\begin{equation} \label{eq:a}
    \mathbf{a} = -\mathbf{p}_{\mathrm{vf}} - \mathbf{p}_{\mathrm{rf}}t_{\mathrm{go}} + \mathbf{p}\frac{t_{\mathrm{go}}^2}{2}
\end{equation}

By substituting (\ref{eq:a}) in the dynamics and integrating, expressions for position and velocity can be obtained as,
\begin{align}
    \mathbf{r} &= \mathbf{r}_f - \mathbf{v}_ft_{\mathrm{go}} - \left( \mathbf{p}_{\mathrm{vf}} - \mathbf{g} \right)\frac{t_{\mathrm{go}}^2}{2} - \mathbf{p}_{\mathrm{rf}}\frac{t_{\mathrm{go}}^3}{6} + \mathbf{p}\frac{t_{\mathrm{go}}^4}{24} \label{eq:r}\\
    \mathbf{v} &= \mathbf{v}_f + \left( \mathbf{p}_{\mathbf{vf}} - \mathbf{g} \right)t_{\mathrm{go}} + \mathbf{p}_{\mathrm{rf}}\frac{t_{\mathrm{go}}^2}{2} - \mathbf{p}\frac{t_{\mathrm{go}}^3}{6} \label{eq:v}
\end{align}
Solving for terminal co-states from (\ref{eq:r}) and (\ref{eq:v}):
\begin{align}
    \mathbf{p}_{\mathrm{rf}} &= \frac{12}{t_{\mathrm{go}}^3} \left[ \left( \mathbf{r} - \mathbf{r}_f \right) + \left( \mathbf{v} + \mathbf{v}_f \right)\frac{t_{\mathrm{go}}}{2} + \mathbf{p}\frac{t_{\mathrm{go}}^4}{24} \right] \label{eq:prf}\\
    \mathbf{p}_{\mathrm{vf}} &= -\frac{6}{t_{\mathrm{go}}^2} \left[ \left( \mathbf{r} - \mathbf{r}_f \right) + \left( \mathbf{v} + 2\mathbf{v}_f \right)\frac{t_{\mathrm{go}}}{3} + \mathbf{p}\frac{t_{\mathrm{go}}^4}{72} \right] + \mathbf{g} \label{eq:pvf}
\end{align}
Finally, substituting (\ref{eq:prf}) and (\ref{eq:pvf}) in (\ref{eq:a}), we get the optimal guidance command as:
\begin{equation}\label{eq:aOptimal}
    \mathbf{a}_{\mathrm{OTALG}} = \frac{6}{t_{\mathrm{go}}^2}\mathbf{ZEM}(t) - \frac{2}{t_{\mathrm{go}}}\mathbf{ZEV}(t) + \mathbf{p}\frac{t_{\mathrm{go}}^2}{12}
\end{equation}
Comparing \eqref{eq:classical_acc} and \eqref{eq:aOptimal}, note that the term $\mathbf{p}(t_{\mathrm{go}}^2/12)$ is responsible for the collision avoidance divert manoeuvre.

\subsection{Discussions}\label{subsec:Discussions}
Define a vector $\mathbf{\Gamma} \triangleq [\Dot{p}_{rx}(t_{\mathrm{go}}^2)/12,\, \Dot{p}_{ry}(t_{\mathrm{go}}^2)/12,\, \Dot{p}_{rz}(t_{\mathrm{go}}^2)/12]^\mathrm{T}$, which is the collision avoidance term in (\ref{eq:aOptimal}). Taking partial derivative of $\Gamma_i$ w.r.t. $r_i$, 
\begin{equation}\label{eq:dGamma}
    \frac{\partial \Gamma_i}{\partial r_i} = \frac{\kappa}{(d_i^2 + l_{1,i})^2} \left[ 1 - \frac{4d_i^2}{(d_i^2 + l_{1,i})} + \frac{2l_{2,i}d_i^2}{(d_i^2 + l_{1,i})^2} \right]e^{-\phi_i}
\end{equation}
where $\kappa = (l_{2,i}l_{3,i}t_{\mathrm{go}}^2)/12$. The non-trivial critical points of (\ref{eq:dGamma}) are when $e^{-\phi_i} = 0$, or $1 - \dfrac{4d_i^2}{(d_i^2 + l_{1,i})} + \dfrac{2l_{2,i}d_i^2}{(d_i^2 + l_{1,i})^2} = 0$. Considering the structure of $\phi_i$, the former does not have real solutions, so only the latter condition needs to be checked to determine the critical points. Simplifying the latter condition, we get the critical values of $d_i$ as:
\begin{align} \label{eq:d_crit}
    d_i^* = \pm\frac{\sqrt{\sqrt{l_{2,i}^2 - 2l_{1,i}l_{2,i} + 4l_{1,i}^2} + l_{2,i} - l_{1,i}}}{\sqrt{3}}.
\end{align}
The maximum magnitude of divert acceleration occurs at $d_i=d_i^*$, which is the critical distance of the barrier to the lander. The safety margin $\delta$ must be chosen to be greater than $d_i^*$. The safety margin can then be tuned according to (\ref{eq:d_crit}) by setting $l_{1,i}$, and $l_{2,i}$ suitably. However, the safety margin of any step should be less than its height. 

From a design perspective, knowledge of the maximum thrust the guidance law needs is a prerequisite. We look at the performance index (\ref{eq:newJ}) to determine $T_{i_{\mathrm{max}}}$ where $i = x,\, y,\, z$. To ensure $J > 0$, the following must hold:
\begin{align}
    a_{i_{\mathrm{max}}}^2 - l_{3,i}\mathrm{e}^{-\phi_i} \geq 0 \Rightarrow |a_{i_{\mathrm{max}}}| \geq \sqrt{l_{3,i}\mathrm{e}^{-\phi_i}} \label{eq:amax}
\end{align}
From (\ref{eq:amax}), we get:
\begin{equation}
    |T_{i_{\mathrm{max}}}| \geq m_0\sqrt{l_{3,i}\mathrm{e}^{-\phi_i}} \label{eq:Tmax}
\end{equation}
From (\ref{eq: p_i}) and (\ref{eq:aOptimal}) $l_{2,i}$ and $l_{3,i}$ have a direct affect on the maximum magnitude of divert acceleration. Thus, increasing $l_{2,i}$ and $l_{3,i}$ require a larger thrust to be generated by the lander, and therefore require a larger $|T_{i_{\mathrm{max}}}|$.

From \eqref{eq:rho_lat}, \eqref{eq:rho_z}, Fig. \ref{fig:barrier} and the definition of $d_i$, we have $\sign r_i = - \sign d_i$. To avoid terrain successfully $|d_i|$ should increase which implies, from \eqref{eq: p_i}, $\sign p_i = \sign d_i$ and therefore the product $l_{2,i}l_{3,i} > 0$. However, from (\ref{eq:amax}), $l_{3,i}>0$ and $l_{2,i} > 0$.

\begin{figure*}
    \centering
    \begin{subfigure}{0.24\textwidth}
        \centering
        \includegraphics[width=\linewidth]{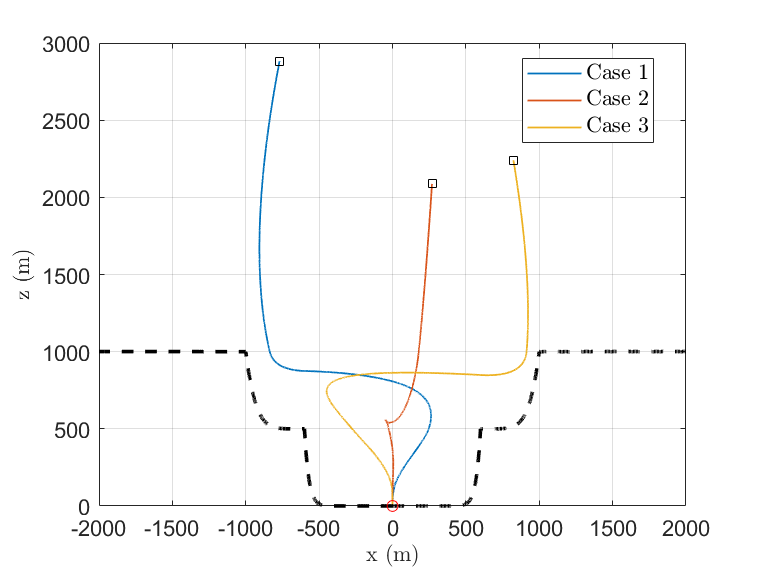} 
        \subcaption{Trajectories in $x$-$z$ plane.}
        \label{fig:xz}
    \end{subfigure}
    \begin{subfigure}{0.24\textwidth}
        \centering
        \includegraphics[width=\linewidth]{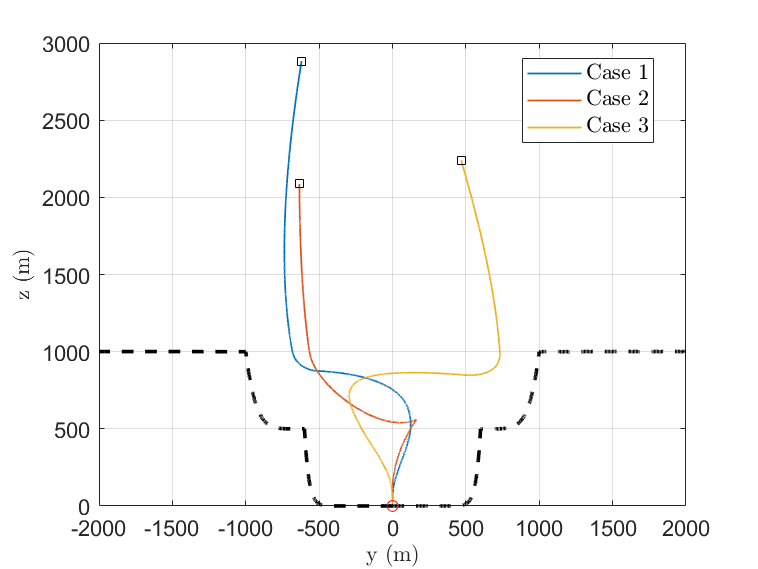} 
        \subcaption{Trajectories in $y$-$z$ plane.}
        \label{fig:yz}
    \end{subfigure}
    \begin{subfigure}{0.24\textwidth}
        \centering
        \includegraphics[width=\linewidth]{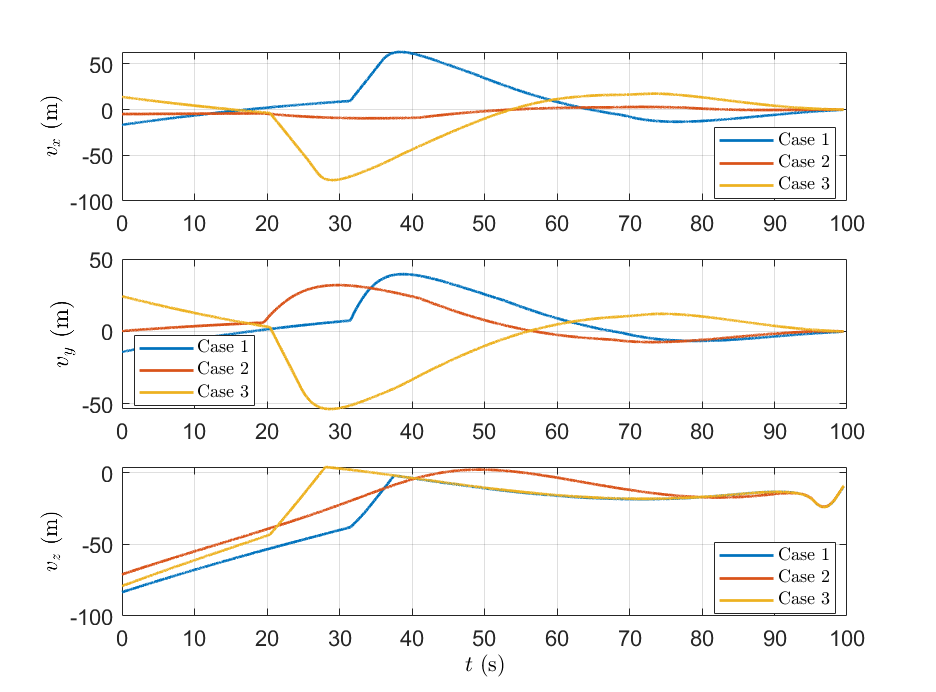}  
        \subcaption{Velocity profiles}
        \label{fig:vel_prof}
    \end{subfigure}
    \begin{subfigure}{0.24\textwidth}
        \centering
        \includegraphics[width=\linewidth]{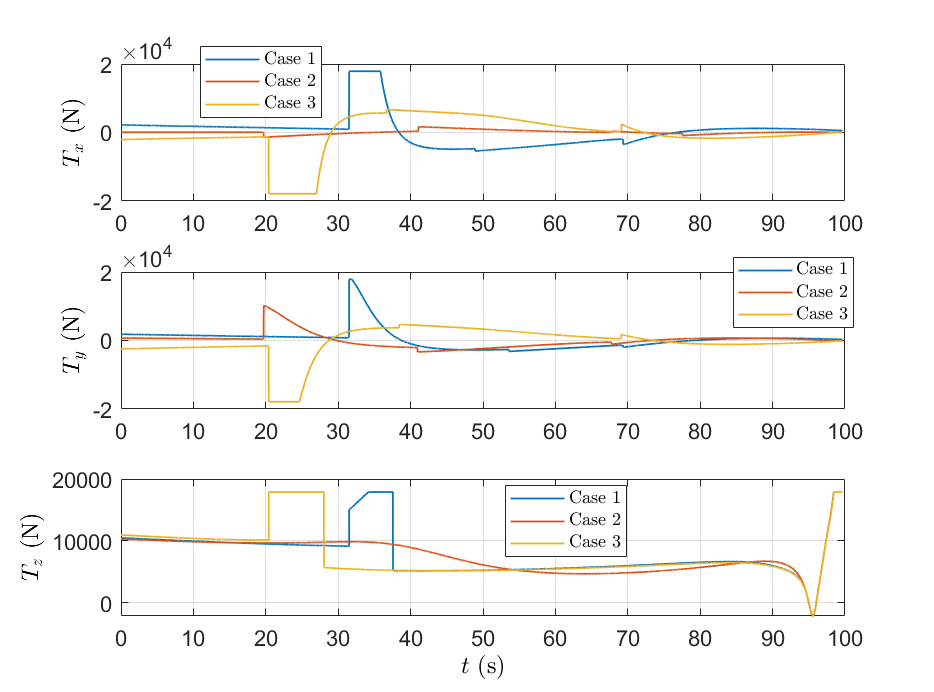} 
        \subcaption{Thrust profiles.}
        \label{fig:thrust}
    \end{subfigure}
    \newline
    \centering
    \begin{subfigure}{0.45\textwidth}
        \centering
        \includegraphics[width=0.8\linewidth]{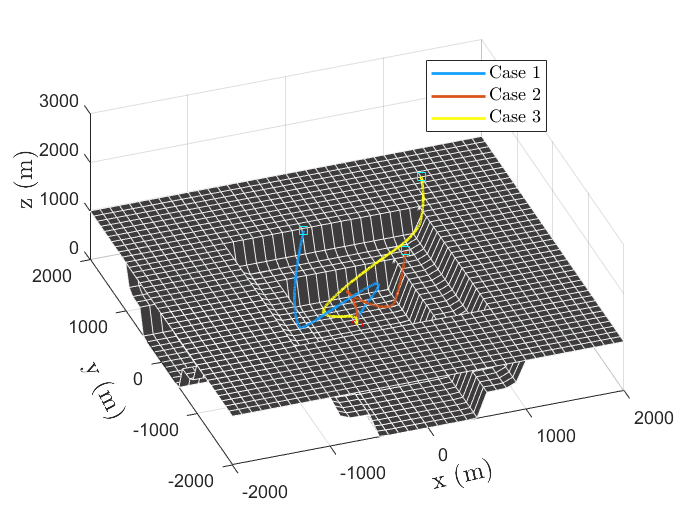} 
        \subcaption{Trajectories in 3D representation - 1.}
        \label{fig:3d_fig1}
    \end{subfigure}
    \begin{subfigure}{0.45\textwidth}
        \centering
        \includegraphics[width=0.8\linewidth]{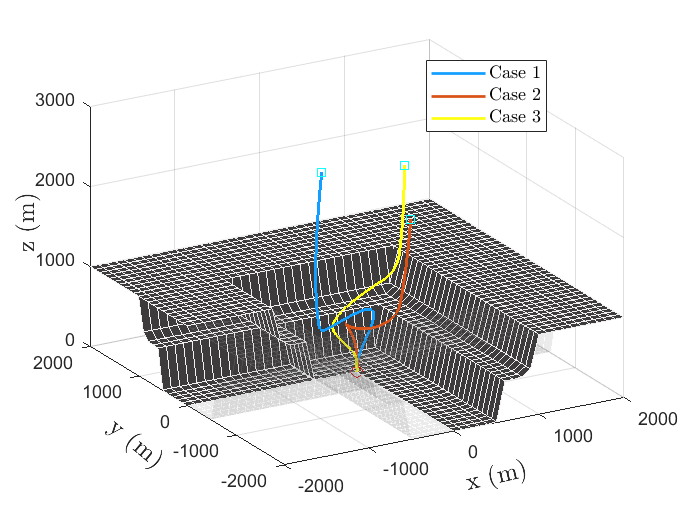} 
        \subcaption{Trajectories in 3D representation - 2.}
        \label{fig:3d_fig2}
    \end{subfigure}
    \label{fig:traj_vel}
    \caption{Trajectories, velocity and thrust profile with no perturbations.}
\end{figure*}

\section{Simulations}\label{sec:Sims}

To demonstrate the effectiveness of the proposed guidance law, we present results from computer simulations in this section. A simplified version of simulation parameters of \cite{Acikmese_Ploen_2007} have been considered for simulation. We assume a point-mass lander with specific impulse $I_{\mathrm{sp}} = 225$ s and $T_{\mathrm{max}} = 31000$ N (equivalent to ten thrusters, each with $3100$ N max thrust). The desired terminal states are $\mathbf{r}_f = [0,\, 0,\, 0]^\mathrm{T}$ m, $\mathbf{v}_f = [0,\, 0,\, 0]^\mathrm{T}$ m/s, at terminal time $t_f = 100$ s. The simulation is stopped when $r_z = 0.05$ m or desired terminal time is achieved, whichever occurs first. To emulate a trench surrounding the landing site on Mars, we consider the terrain that can be modelled as a $2$-step, flat-top shape. The height and width from the origin of each step are given in Table \ref{tab:terrain}. To design the barriers (see \eqref{eq:rho_lat} and \eqref{eq:rho_z}), we  choose $\theta_{3} = 0.05^{\circ}$, with $\lambda_{i,2} = 6$, and $\lambda_{i,1} = 20$. Further, the margin of safety in the vertical barrier is $\delta = 1.2d_{i}^*$. The guidance law constant are chosen as $l_{1,i} = 1,\ l_{2,i} = 3000,\ $ and $l_{3,i} = 280$. Thus, from \eqref{eq:d_crit}, we have $\delta = 53.67$ m. Finally, the local gravity at Mars is assumed to be $g = [0,\ 0,\, -3.7114]^\mathrm{T}$ m/s\textsuperscript{2}, and acceleration due to gravity on Earth $g_e = 9.807$ m/s\textsuperscript{2}.

\begin{table}[h]
\begin{center}
\caption{Model for terrain approximation}\label{tab:terrain}
\begin{tabular}{lcc}
 & Height (m) & Width (m)\\
 \hline
 Step-$1$ & $500$ & $600$\\
 Step-$2$ & $1000$ & $1000$\\
\end{tabular}
\end{center}
\end{table}

\begin{table}
\begin{center}
\caption{Sample Initial Conditions for Illustration}\label{tab:ic_3}
\begin{tabular}{l c c c}
\hline Case \# & $\mathbf{r}_0$ & $\mathbf{v}_0$ & $m_0$\\
    \hline
    \multirow{2}{*}{Case 1} & $[-769.42, -619.63,$ & $[-16.78, -14.08,$ &\multirow{3}{*}{1961.80}\\
     & $2883.33]^\mathrm{T}$ & $-83.36]^\mathrm{T}$ & \\
     \multirow{2}{*}{Case 2} & $[269.35, -634.30, $ & $[-4.98, 0.29, $ &\multirow{3}{*}{1916.55}\\
     & $2086.65]^\mathrm{T}$ & $-70.89]^\mathrm{T}$ & \\
     \multirow{2}{*}{Case 3} & $[823.91, 467.70, $ & $[13.81, 24.28, $ &\multirow{3}{*}{1959.43}\\
     & $2240.03]^\mathrm{T}$ & $-79.47]^\mathrm{T}$ & \\
\end{tabular}
\end{center}
\end{table}

\subsection{Terrain collision avoidance}\label{subsec:TerrCollAvoid}

To portray the effectiveness of the proposed guidance law in terms of terrain avoidance, we present the trajectories of $N = 3$ sample initial conditions. Trajectories corresponding to these initial conditions are shown in Fig. \ref{fig:xz}, Fig. \ref{fig:yz}, Fig. \ref{fig:3d_fig1} and Fig. \ref{fig:3d_fig2}. The velocity profiles of the three cases are shown in Fig. \ref{fig:vel_prof}, and the thrust profiles are shown in Fig. \ref{fig:thrust}.

\begin{table}
\begin{center}
\caption{Initial Conditions Distribution Setup}\label{tab:ic}
\begin{tabular}{l c c p{3cm}}
\hline Initial Condition & $\mu$ & $\sigma^2$\\
    \hline
    $x_0$ (m) & $0$ &  350\\
    $y_0$ (m) & $0$ & 350\\
    $z_0$ (m) & $2500$ & $500$\\
    $v_{\mathrm{x0}}$ (m/s) & $0$ & 10\\
    $v_{\mathrm{y0}}$ (m/s) & $0$ & 10\\
    $v_{\mathrm{z0}}$ (m/s) & $-80$ & $10$\\
    $m_0$ (kg) & $1900,\, 2100$ & 100\\
    \hline
\end{tabular}
\end{center}
\end{table}

It is clear from the velocity profile in Fig. \ref{fig:vel_prof} that the guidance law brings the lander softly to the landing site. There are no sudden jerks in either the trajectories or the velocity profiles. We can observe the guidance law working to avoid the terrain from the trajectory figures and the velocity profiles. Consider Case 3 in Fig. \ref{fig:xz}. When the lander is near the terrain, the acceleration command generated, as shown in Fig. \ref{fig:thrust}, by the collision avoidance term pushes it away from the terrain by changing the velocity direction, as clearly observed in the $v_x$ plot of Fig. \ref{fig:vel_prof}. In all three cases as the lander approaches landing site, the proposed guidance law generates the acceleration in the direction opposite to the direction of motion as evident from Fig. \ref{fig:thrust}, and softly reaches the landing site.

\subsection{Fuel consumption}\label{subsec:FuelConsmp}

\begin{figure*}
    \centering
    \begin{subfigure}{0.45\textwidth}
        \includegraphics[width=0.7\linewidth]{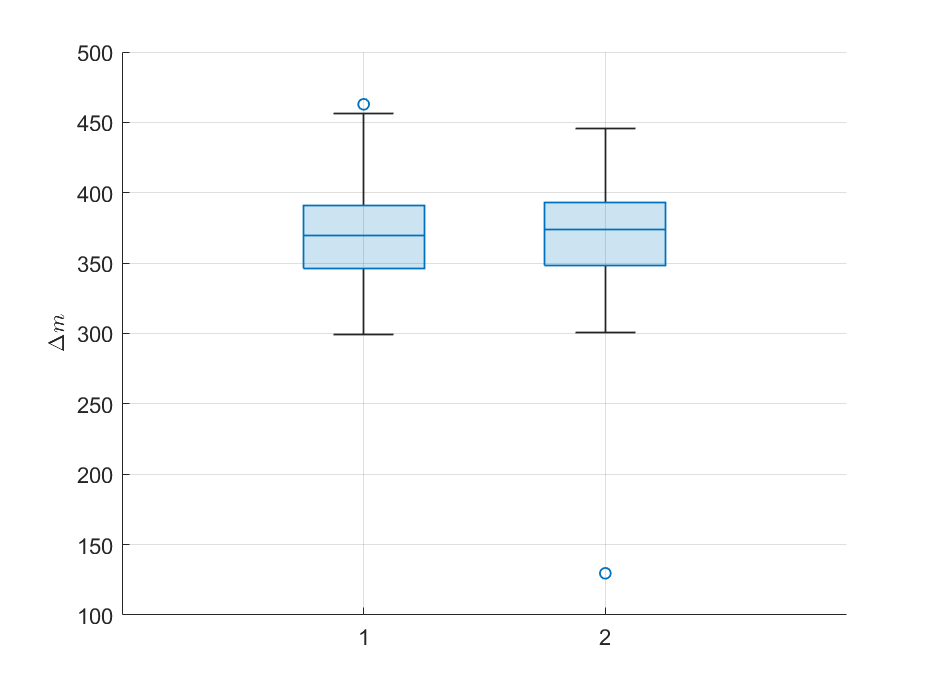}
        \subcaption{Fuel consumption statistics under no perturbations.}
        \label{fig:box}
    \end{subfigure}
    \begin{subfigure}{0.45\textwidth}
        \includegraphics[width=0.7\linewidth]{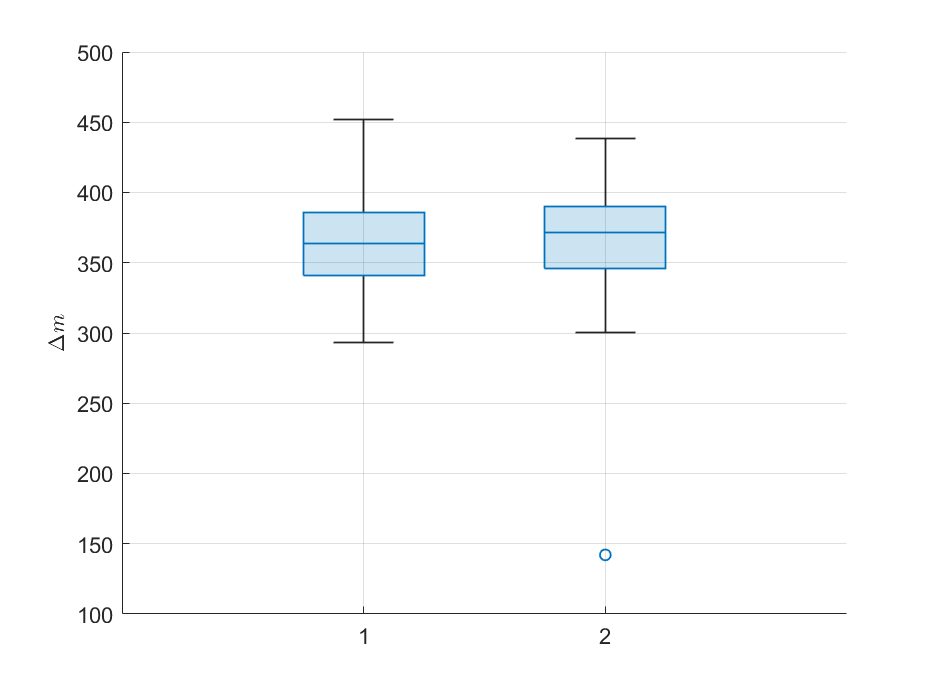}  
        \subcaption{Fuel consumption statistics under with atmospheric perturbations.}
        \label{fig:box_ap}
    \end{subfigure}
    \caption{Fuel consumption statistics.}
    \label{fig:stats}
\end{figure*}

To compare the efficacy of the proposed guidance law (Algorithm 1) in terms of fuel consumption, we compare it with the work done in \cite{Zhang_Guo_Ma_Zeng_2017} (Algorithm 2), which was shown to be near-fuel optimal with respect to the classical ZEM/ZEV guidance law in \cite{Ebrahimi_Bahrami_Roshanian_2008}. Further, the performance is also tested under bounded atmospheric perturbations modelled as
\begin{equation} \label{eq:ap}
    \mathbf{a}_p(t) = 0.5\mathbf{a}_\mathrm{OTALG}\sin{\left( \frac{\pi}{3}t \right)}.
\end{equation}
where, $\mathbf{a}_\mathrm{OTALG}$ is as obtained in \eqref{eq:aOptimal}. We perform paired t-test between the net fuel consumption of the two guidance laws. In this test, there are $N = 300$ data points for fuel consumption, initial conditions for which are taken from a normally distributed set in Table \ref{tab:ic}. The null hypothesis for the test is $H_0: d\triangleq\Delta m_1 - \Delta m_2=0$, where $\Delta m_j$ is the fuel consumption for Algorithm $j$. For $0.05$ significance level, we have $t_{\mathrm{crit}} =1.9679$. We find $\Bar{d} = -0.3581$ and $\sigma_{\Bar{d}} = 1.2719$. Thus, for the paired t-test, $t = -0.2815$ implying $|t|<t_{\mathrm{crit}}$. From the box chart in Fig. \ref{fig:box}, we find that the median fuel consumption is marginally smaller for the proposed law. Thus, we can infer that the proposed guidance law w.r.t. that in \cite{Zhang_Guo_Ma_Zeng_2017} are statistically almost similar in terms of near-fuel optimal performance under ideal scenario. However, as We perform the paired t-test on the data-set generated with the same initial conditions under atmospheric perturbations as defined in \eqref{eq:ap}. We get $\Bar{d} = -4.2582$ and $\sigma_{\Bar{d}} = 1.2116$. Thus, $t = -3.5415$ implying that $|t|>t_{\mathrm{crit}}$. Moreover, from the box charts in Figure \ref{fig:box_ap} for the net fuel consumption under atmospheric perturbations, it is observed that the median fuel consumption is lesser in Algorithm 1. With these, we can infer that under bounded atmospheric perturbations, the proposed guidance law achieves far superior performance in terms of near-fuel optimality when compared to Algorithm 2. 

\section{Conclusion}\label{sec:Concl}

Near-Optimal guidance law for landing in hazardous terrain with minimal fuel consumption has been presented in this paper. The terrain has been modelled as $n$-step-shaped polygons, and barriers have been defined over the terrain model using piece-wise smooth polynomials. The efficacy of the guidance law is tested in a trench in the Martian environment, emulated using a flat-top, 2-step shape. The developed guidance algorithm successfully achieves precision soft landing, while also avoiding any collision with the terrain. Using extensive simulations, the near-fuel optimality is confirmed. The guidance law avoids terrain under bounded perturbation and maintains near-fuel optimality. Study on barrier profiles and enhancing the robustness of the spacecraft landing guidance form a future scope of research.
\bibliography{ifacconf}{}
\end{document}